\documentclass[prd,onecolumn,notitlepage,eqsecnum,nofootinbib,floatfix]{revtex4-1}
\usepackage{amssymb}
\usepackage{amsmath}
\usepackage{amsfonts} 
\usepackage{bm}
\usepackage{graphicx}
\usepackage{xcolor}
\allowdisplaybreaks
\begin{document}
\title{Swimming in spacetime: the view from a Fermi observer} 
\author{Raissa F.\ P.\ Mendes}  
\affiliation{Instituto de F\'isica, Universidade Federal Fluminense, \\
Av.~Gal.~Milton Tavares de Souza s/n, Gragoat\'a, 24210-346 Niter\'oi, 
Rio de Janeiro, Brazil.}
\author{Eric Poisson} 
\affiliation{Department of Physics, University of Guelph, Guelph,
  Ontario, N1G 2W1, Canada}
\date{\today} 

\begin{abstract} 
An extended test body moving in a curved spacetime does not typically
follow a geodesic, because of forces that arise from couplings between
its multipole moments and the ambient curvature. An illustration of this
fact was provided by Wisdom, who showed that the motion of a quasi-rigid
body undergoing cyclic changes of shape in a curved spacetime deviates, 
in general, from a geodesic. Wisdom's analysis, however, was
recently challenged on the grounds that the body's motion should be
described by the Mathisson-Papapetrou-Dixon equations, and that these
predict geodesic motion for the kind of body considered by Wisdom. We
attempt to shed some light on this matter by examining the motion of
an internally-moving tripod in Schwarzschild spacetime, as viewed by a
Fermi observer moving on a timelike geodesic. We find that the
description of the motion depends sensitively on a choice of cycle for
the tripod's internal motions, but also on a choice of  ``center of
mass'' for the tripod; a sensible (though not unique) prescription for
this ``center of mass'' produces a motion that conforms with Wisdom's
prediction: the tripod drifts away from the observer, even when they
are given identical initial conditions. We suggest pathways of
reconciliation between this conclusion and the null result that
apparently follows from the Mathisson-Papapetrou-Dixon equations of
motion.     
\end{abstract}

\maketitle

\section{Introduction} 
\label{sec:intro} 

\noindent
\textit{
[Disclaimer: In this work we employ a constrained Lagrangian formalism 
in order to model a body undergoing cyclic changes of shape in a curved
spacetime. However, this constrained Lagrangian formalism is not 
necessarily consistent with motions generated by internal forces alone,
and may exhibit undesirable features even in special relativity.
For instance, from the discussion in Sec.~\ref{subsec:flat}
it can be seen that if a body initially at rest with respect to some inertial 
observer starts a cyclic motion, it may acquire a nonzero velocity 
with respect to the same observer for some choices of cycles.
It appears, therefore, that additional restrictions should be imposed
on the cycle for it to be consistent with the operation of internal 
forces alone, otherwise we might be led to interpret as ``swimming'' 
something that is due to an unreasonable power of the agent 
responsible for keeping the body's motion. 
Considerations along these lines, raised largely through a discussion with 
Amos Ori, cast some doubts on the applicability of the constrained Lagrangian
formalism to study swimming in curved spacetimes. 
Nonetheless, we still believe the calculations presented here are technically
correct, and might be instructive for future research on related problems.]
}
\\

It is a well-known fact that in general relativity, an extended 
test body may not move on a timelike geodesic, because of forces that
typically arise from couplings between its multipole moments and the 
ambient curvature.
Wisdom provided a vivid illustration of this fact in a 2003 paper 
\cite{wisdom:03}, where he demonstrated that a test 
body undergoing  cyclic changes of shape --- a swimmer --- does not 
move on  a geodesic. Wisdom gave a concrete example of this effect 
by calculating the displacement of a tripod relative to a geodesic 
in Schwarzschild spacetime; he found the displacement to scale with 
the spacetime curvature and the extent of the tripod's internal
motions. The effect was shown to be small and uninteresting 
from a practical point of view, but it is nevertheless
interesting as a matter of principle.   

The validity of Wisdom's analysis, however, was recently called to 
question by Silva, Matsas, and Vanzella
\cite{silva-matsas-vanzella:16}. The objection raised in their work 
relies on the observation that the swimmer's motion ought
to be governed by the Mathisson-Papapetrou-Dixon (MPD) equations
\cite{mathisson:10, papapetrou:51a, dixon:70a, dixon:70b, dixon:74},  
\begin{equation} 
\frac{Dp_\alpha}{d\tau} = \frac{1}{2} S^{\mu\nu} u^\lambda
R_{\mu\nu\lambda\alpha}, \qquad 
\frac{DS_{\alpha\beta}}{d\tau} = 2 p_{[\alpha} u_{\beta]},  
\label{extended_motion} 
\end{equation}  
where $p_\alpha$ is the body's momentum vector, $S_{\alpha\beta}$ its
spin tensor, $u^\alpha$ the tangent to the world line, 
$R_{\alpha\beta\gamma\delta}$ the Riemann tensor, and $D/d\tau$
indicates covariant differentiation with respect to proper time
$\tau$. These equations are meant to describe the motion of a generic
extended body, in a pole-dipole approximation that neglects the
influence of higher multipole moments. The MPD equations
indicate that the force acting on an extended body should scale with 
the curvature, as was observed in Ref.~\cite{wisdom:03}, but after a deeper
examination, the authors of Ref.~\cite{silva-matsas-vanzella:16} conclude 
that they are incompatible with the swimming motion revealed by Wisdom; 
according to Eq.~(\ref{extended_motion}), the tripod should move on a
geodesic. The authors attempt to rescue the phenomenon by suggesting 
that the motion might scale with the covariant derivative of the
Riemann tensor, instead of the Riemann tensor itself, but this
suggestion is incompatible with Wisdom's findings.   

We take this opportunity to revisit Wisdom's original analysis, and to 
calculate in a novel way the motion of an internally-moving tripod in
Schwarzschild spacetime. Instead of describing the motion in the
static frame of the Schwarzschild spacetime, as Wisdom did, we
prefer to exploit a reference frame attached to a freely moving
observer who follows a timelike geodesic. Observer and tripod 
are given identical initial positions and velocities in the 
spacetime, and the tripod's motion is measured relative to the 
observer's rest frame. Our implementation of this idea relies on 
Fermi normal coordinates, which allow us to write the metric in a 
convenient locally-flat form. The Riemann tensor appears explicitly in the
metric, and this clarifies its influence on the tripod's motion. And
because the Fermi coordinates are constructed from geodesic segments
that are everywhere orthogonal to the observer's world line, they come
with a transparent geometrical meaning that helps clarify
the description of the motion.     

We begin in Sec.~\ref{sec:fermi} with a review of Fermi normal
coordinates attached to a radial, timelike geodesic in a static,
spherically symmetric spacetime. In Sec.~\ref{sec:tripod1} we
formulate our precise model for the tripod, specify the cycle of its 
internal motions, construct its Lagrangian, and derive the equations  
of motion. A delicate matter that presents itself is the designation
of an appropriate ``center of mass'' (CM) for the tripod, which we use
to track the motion of the tripod as a whole. Our relativistic
definition is based on the requirements that there
should be no swimming in Minkowski spacetime, and that 
the CM should always be contained within the body for any cycle of
internal motions. These requirements determine the CM position up
to a constant shift, and completing the prescription with a 
specification of this shift turns out to be an important aspect of our 
analysis, with a significant bearing on our conclusions.  

We integrate the equations of motion in Sec.~\ref{sec:tripod2}. 
We begin with a discussion of motion in de Sitter spacetime, and show
that the shift freedom in the CM definition can be exploited to remove
a drift from geodesic motion that would otherwise be present. With the
CM position fully specified by this prescription, we place the tripod
in the Schwarzschild spacetime and show that the coupling between
its internal motions and the spacetime curvature prevents it from
following a geodesic. The tripod's world line is not a geodesic, but
we find that it asymptotes to a geodesic when the tripod is allowed to 
approach the central singularity of the Schwarzschild metric; the
asymptotic geodesic is distinct from the reference geodesic of the
freely-falling observer.  

In Sec.~\ref{sec:MPD} we propose two paths of reconciliation with 
the MPD equations. In the first, we suggest that the MPD equations may
not apply to the tripod, because their derivation relies crucially on 
energy-momentum conservation. The tripod, on the other hand, does not
conserve energy and momentum, because external agents are
required to keep the tripod on its cycle, and these can supply the
missing energy and momentum. We provide evidence to support this
suggestion by generalizing the MPD equations to a constrained
mechanical system, and showing that the resulting equations do differ
from Eqs.~(\ref{extended_motion}). The second path of reconciliation
is a suggestion that the apparent incompatibility between Wisdom's
swimming and the MPD equations might not be real, but the result of a 
misinterpretation of the equations. The main point is that
Eqs.~(\ref{extended_motion}) are empty of content until a relation
between $p_\alpha$ and $u_\alpha$ is specified through the selection
of a suitable ``center of mass'' for the extended body. Our
considerations in Sec.~\ref{sec:tripod1} remind us that this can be a
very delicate matter. A simple extended body might motivate the
introduction of a simple auxiliary condition such as $p_\alpha
S^{\alpha\beta} = 0$, which yields an explicit relation between
$p_\alpha$ and $u_\alpha$. But a tripod undergoing cyclical internal
motions is not simple, and it is likely that the relation between
$p_\alpha$ and $u_\alpha$ is far more complicated in this case,
involving the details of the tripod's design. And until this relation
is identified, the predictions of Eq.~(\ref{extended_motion}) must
remain ambiguous.      

\section{Fermi coordinates} 
\label{sec:fermi} 

The motion of an extended body in any spacetime can be described from
the point of view of an observer moving on a reference timelike
geodesic $\gamma$, which is imagined to stay within the body's
neighborhood. The coordinates that best capture this point
of view are the Fermi normal coordinates $(t, x^a)$, in which $t$ 
measures proper time on $\gamma$, and the spatial coordinates $x^a$
measure proper distance on spacelike geodesics that are orthogonal to
$\gamma$. In relativistic units with $c=1$, used throughout the paper,
the metric of the spacetime is given by     
\begin{subequations} 
\label{Fermi_metric} 
\begin{align} 
g_{00} &= -1 + h_{00} = -1 - R_{0a0b}(t)\, x^a x^b, \\ 
g_{0a} &= h_{0a} = -\frac{2}{3} R_{0bac}(t)\, x^b x^c, \\ 
g_{ab} &= \delta_{ab} + h_{ab} = \delta_{ab} 
- \frac{1}{3} R_{acbd}(t)\, x^c x^d,  
\end{align} 
\end{subequations} 
up to cubic terms in the spatial coordinates; the components of the
Riemann tensor are evaluated on $\gamma$ and expressed as functions of
proper time $t$. The metric is flat on $\gamma$, with a 
vanishing connection, and the Fermi coordinates define the rest 
frame of the observer moving on $\gamma$. (An introduction to Fermi 
coordinates can be found in Ref.~\cite{manasse-misner:63}. An 
alternative is Sec.~9 of Ref.~\cite{poisson-pound-vega:11}.)    

We consider a static, spherically symmetric spacetime with metric  
\begin{equation} \label{metric}
ds^2 = -f\, dT^2 + f^{-1}\, dR^2 + R^2 (d\Theta^2 
+ \sin^2\Theta\, d\Phi^2), 
\end{equation} 
in which $f$ is a function of $R$. For concreteness below we shall 
take $f = 1 - 2GM/R$, so that the metric is that of the
Schwarzschild spacetime, which describes the geometry outside a spherical
body of mass $M$. We will also consider the case of de Sitter 
spacetime, for which $f = 1 - k^2 R^2$, with $k^2$ proportional to the  
spacetime curvature. The reference timelike geodesic
$\gamma$ is chosen to be a radial world line with tangent vector   
\begin{equation} 
u^\alpha = \bigl[ u^T, u^R, u^\Theta, u^\Phi \bigr] 
= \Bigl[ E/f, -\sqrt{E^2-f}, 0, 0 \Bigr], 
\end{equation} 
where $E$, the dimensionless energy parameter, is a constant of the
motion. The radial component $u^R$ of the tangent vector is negative,
which indicates that $R$ decreases along $\gamma$; in the context of
the Schwarzschild spacetime, the reference observer falls toward the
central object of mass $M$.  

The vectors $e^\alpha_a$ are defined to be mutually orthogonal,
orthogonal to $u^\alpha$, and parallel transported on $\gamma$. It is
easy to show that the set  
\begin{subequations} 
\label{triad} 
\begin{align} 
e^\alpha_1 &= \bigl[ 0, 0, 0, -1/(R\sin\theta) \bigr], \\ 
e^\alpha_2 &= \bigl[ 0, 0, 1/R, 0 \bigr], \\ 
e^\alpha_3 &= \bigl[ -\sqrt{E^2-f}/f, E, 0, 0 \bigr] 
\end{align} 
\end{subequations} 
satisfies these requirements. The vector $e^\alpha_1$ points in the
direction of decreasing $\Phi$, $e^\alpha_2$ in the direction of
increasing $\Theta$, and $e^\alpha_3$ is mostly aligned with the
direction of increasing $R$. The triad forms a right-handed system,
and the third direction is identified with the ``up'' direction. 

The components of the Riemann tensor in Fermi coordinates are equal to
its projections in the tetrad formed by $u^\alpha$ and
$e^\alpha_a$. We have 
\begin{equation} 
R_{0a0b} = R_{\mu\alpha\nu\beta}\, u^\mu e^\alpha_a u^\nu e^\beta_b,
\qquad 
R_{0bac} = R_{\mu\beta\alpha\gamma}\, u^\mu e^\beta_b e^\alpha_a
e^\gamma_c, \qquad 
R_{acbd} = R_{\alpha\gamma\beta\delta}\, e^\alpha_a e^\gamma_c
e^\beta_b e^\delta_d, 
\end{equation} 
and calculation yields the nonvanishing components 
\begin{equation} 
R_{0101} = R_{0202} = - R_{1313} = -R_{2323} = A, \qquad 
R_{0303} = B, \qquad 
R_{1212} = C, 
\label{Riemann_components} 
\end{equation} 
where 
\begin{equation} 
A := \frac{f'}{2R}, \qquad 
B  := \frac{1}{2} f'', \qquad 
C := \frac{1-f}{R^2},  
\label{ABCdef} 
\end{equation} 
with a prime indicating differentiation with respect to $R$. Making
the substitutions in Eq.~(\ref{Fermi_metric}) produces the
nonvanishing components of the metric perturbation $h_{\alpha\beta}$. 
In particular, $h_{0a} = 0$ for our choice of $\gamma$.  

\section{Tripod model}  
\label{sec:tripod1}      

We wish to determine the motion of a swimmer in the spacetime 
described in Sec.~\ref{sec:fermi}. We adopt Wisdom's model 
\cite{wisdom:03}, in which the swimmer is given the shape of a tripod;
refer to Fig.~3 of his paper. The tripod consists of a ``head'' of
mass $m_0$ and three ``feet'' of equal mass $m_1$. The feet are
attached to the head with massless struts. Each strut has a length
$\ell(t)$ and makes an angle $\alpha(t)$ with the axis of symmetry. 
We construct the tripod's Lagrangian by adding the individual
Lagrangians of the head and feet and incorporating the constraints
enforced by the struts; we neglect the stresses in the struts.   

\subsection{Newtonian description} 

We begin with a Newtonian description of the tripod, in the absence of
gravity. In an inertial frame $(x, y, z)$, the coordinates of the
tripod's head are denoted $\bm{r}_0 = (x_0, y_0, z_0)$. The
coordinates of each foot are given by 
\begin{subequations} 
\label{constraints} 
\begin{align}    
\bm{r}_1 - \bm{r}_0 &= \biggl( \frac{\sqrt{3}}{2} \ell \sin\alpha, 
-\frac{1}{2} \ell \sin\alpha, -\ell \cos\alpha \biggr), \\  
\bm{r}_2 - \bm{r}_0 &= (0, \ell\sin\alpha, -\ell\cos\alpha), \\ 
\bm{r}_3 - \bm{r}_0 &= \biggl( -\frac{\sqrt{3}}{2} \ell \sin\alpha, 
-\frac{1}{2} \ell \sin\alpha, -\ell \cos\alpha \biggr). 
\end{align} 
\end{subequations} 
We adopt the position of the tripod's center of mass (CM), given by 
\begin{equation} 
x = x_0, \qquad y = y_0, \qquad 
z = z_0 - \frac{3m_1}{m_0 + 3m_1} \ell \cos\alpha, 
\label{CM} 
\end{equation} 
as generalized coordinates. A simple calculation then reveals that up
to an irrelevant function of time, the tripod's Lagrangian 
$L := L_0 + L_1 + L_2 + L_3$, with 
$L_i = \frac{1}{2} m_i (\dot{x}_i^2 + \dot{y}_i^2 + \dot{z}_i^2)$, 
is given by 
\begin{equation} 
\bar{L} := \frac{L}{m_0 + 3m_1} = \frac{1}{2} (\dot{x}^2 + \dot{y}^2 
+ \dot{z}^2), 
\end{equation} 
in which an overdot indicates differentiation with respect to $t$. 
This Lagrangian is identical to that of a free particle, and we
conclude that the tripod's CM will stay at rest if it begins at
rest. There is no swimming in this Newtonian description.

\subsection{Relativistic description: flat spacetime} 
\label{subsec:flat} 

We continue to ignore gravity, and consider the tripod's dynamics in
special relativity. We actually consider an approximate description
in which all speeds are assumed to be small compared to the speed of
light, so that only the leading relativistic correction is
incorporated in each particle's Lagrangian,  $L_i =
\frac{1}{2} m_i v_i^2 + \frac{1}{8} m_i v_i^4$ with $v_i^2 =
\dot{x}_i^2 + \dot{y}_i^2 + \dot{z}_i^2$. We work in a Lorentz
frame $(t, x, y, z)$ and continue to relate the coordinates of the
feet to those of the head by Eq.~(\ref{constraints}). We make a
relativistic adjustment to the CM variables, so that they are now
given by   
\begin{equation} 
x = x_0, \qquad y = y_0, \qquad 
z = z_0 - \frac{3m_1}{m_0 + 3m_1} \ell \cos\alpha - \delta z, 
\label{CM_flat} 
\end{equation} 
where $\delta z$ shall be determined below.  Making the substitutions
in the tripod's Lagrangian, we observe that it has the structure 
\begin{equation} \label{LMinkowski}
\bar{L} = \frac{1}{2} (\dot{x}^2 + \dot{y}^2 + \dot{z}^2)
+ \frac{1}{8} (\dot{x}^2 + \dot{y}^2 + \dot{z}^2)^2
+ a_1(t) \dot{x}^2 + a_2(t) \dot{y}^2 + a_3(t) \dot{z}^2 
+ a_4(t) \dot{z}, 
\end{equation} 
where each $a_i$ depends on the time derivative of 
$\ell\cos\alpha$ and $\ell\sin\alpha$; the function $a_4(t)$ also  
implicates $\delta\dot{z}$. 

Our prescription\footnote{There would be no need for such a
  prescription if we had access to a complete energy-momentum tensor
  for the tripod. The CM would then be defined in terms of this
  tensor. But we do not have such an object, because the external
  agents responsible for keeping the tripod on its cycle of internal
  motions are not explicitly accounted for in the model. There is
  therefore no energy-momentum tensor, no statement of energy-momentum
  conservation, and no definition of a CM.}  for the CM adjustment
$\delta z$ is based on the requirements that (i) the CM should stay
within the body for any cycle of internal motions, and (ii) the CM
should move uniformly on a straight path; in particular, the CM should
stay at rest if it begins at rest in the adopted Lorentz frame. Now,
the form of the Lagrangian in Eq.~(\ref{LMinkowski}) implies that 
$p_z := \partial{\bar{L}}/\partial \dot{z} = \dot{z}(1 + \cdots) +
a_4(t)$, where the ellipsis represents relativistic corrections; 
$p_z$ is a constant of the motion by virtue of the Euler-Lagrange 
equation. It follows that $p_z = a_4(0)$ when we impose the initial
condition $\dot{z}(0) = 0$, taking the CM to be initially at rest. At
later times we have that $a_4(0) = \dot{z}(1 + \cdots) +
a_4(t)$, and we find that $\dot{z} \neq 0$ unless 
$a_4(t) = a_4(0)$. In other words, a CM initially at rest will
be moving at later times, in violation of our second requirement,
unless we demand that $a_4(t)$ be a constant. Because this function
implicates $\delta\dot{z}$, we find that the condition 
$a_4(t) = a_4(0)$ implies 
\begin{equation} 
\delta \dot{z} = \frac{3 m_0 m_1}{2(m_0 + 3m_1)^2} \Biggl\{  
\frac{m_0-3m_1}{m_0 + 3m_1} 
\biggl[ \frac{d}{dt}(\ell\cos\alpha) \biggr]^2 
+ \biggl[ \frac{d}{dt}(\ell\sin\alpha) \biggr]^2 \Biggl\} 
\frac{d}{dt} (\ell\cos\alpha) + k, 
\label{CMcorr} 
\end{equation} 
where $k := a_4(0)$.

To determine $k$ we appeal to our first requirement. We observe 
that for a given choice of cyclic functions $\ell(t)$ and $\alpha(t)$,
the right-hand side of Eq.~(\ref{CMcorr}) will be a periodic function
with (typically) a nonzero average, giving rise to a $\delta z$ that
features periodic oscillations superposed to a linear growth. To kill
the growth and ensure that the CM does not drift away from the body,
we choose $k$ in such a way that the average of $\delta \dot{z}$
vanishes. In this way, our requirements determine the CM completely,
except for the remaining freedom to choose the initial condition
$\delta z(0)$ when integrating Eq.~(\ref{CMcorr}). With this
prescription, there is no swimming in flat spacetime.  
 
\subsection{Relativistic description: curved spacetime} 

We next incorporate gravity by placing the tripod in the spacetime
described in Sec.~\ref{sec:fermi}.  The coordinates $(x, y, z)$ now
refer to the Fermi frame attached to a radial geodesic of the
spacetime described by Eq.~(\ref{metric}), and $t$ is proper time 
measured by an observer moving on this geodesic.   

A point mass $m$ moving freely in the spacetime is described by the
word line $x^\alpha = r^\alpha(t)$, with $r^0 = t$. The coordinate
velocities are $v^\alpha = dr^\alpha/dt$, with $v^0 = 1$. Taking into
account that $h_{0a} = 0$, the particle's Lagrangian is   
\begin{equation} 
L = -m \bigl( -g_{\alpha\beta} v^\alpha v^\beta \bigr)^{1/2} 
= -m \bigl( 1 - v^2 - h_{00} - h_{ab} v^a v^b \bigr)^{1/2},  
\end{equation} 
in which $v^2 := \delta_{ab} v^a v^b$. As we did previously, we assume
that $v \ll 1$ and expand $L$ in powers of $v^2$, taking
$h_{\alpha\beta}$ to be of order $v^2$ and keeping $L$ linear in the
curvature (that is, neglecting terms quadratic in $h_{00}$). If we
also discard the irrelevant constant term $-m$, the Lagrangian becomes  
\begin{equation} 
L/m = \frac{1}{2} v^2 + \frac{1}{2} h_{00} + \frac{1}{8} v^4  
+ \frac{1}{4} h_{00} v^2 + \frac{1}{2} h_{ab} v^a v^b. 
\label{Lagrangian1} 
\end{equation} 
We recognize the Newtonian kinetic energy $\frac{1}{2} v^2$ and its
relativistic correction $\frac{1}{8} v^4$, the Newtonian potential
energy $\frac{1}{2} h_{00}$, and the remaining terms provide additional
relativistic corrections to the Lagrangian. Substitution of
Eqs.~(\ref{Fermi_metric}) and (\ref{Riemann_components}) into
Eq.~(\ref{Lagrangian1}) gives the explicit expression    
\begin{align} 
L/m  &= \frac{1}{2} {\sf v}^2 - \frac{1}{2} A ({\sf x}^2 + {\sf y}^2) 
- \frac{1}{2} B {\sf z}^2 + \frac{1}{8} {\sf v}^4
\nonumber \\ & \quad \mbox{} 
- \frac{1}{12} A \bigl[ ({\sf x}^2+{\sf y}^2)(3 \dot{\sf x}^2 
   + 3 \dot{\sf y}^2 
  + \dot{\sf z}^2 ) - 2 {\sf z}^2 (\dot{\sf x}^2 + \dot{\sf y}^2) 
  + 4( {\sf x}\dot{\sf x} + {\sf y}\dot{\sf y} ) {\sf z} \dot{\sf z} \bigr] 
\nonumber \\ & \quad \mbox{} 
- \frac{1}{4} B {\sf z}^2 {\sf v}^2 
- \frac{1}{6} C \bigl( {\sf y}^2 \dot{\sf x}^2 - 2 {\sf xy}\, 
    \dot{\sf x} \dot{\sf y} 
  + {\sf x}^2 \dot{\sf y}^2 \bigr), 
\label{Lagrangian2} 
\end{align} 
where $({\sf x, y, z})$ are the components of the spatial vector
$r^a$, $({\sf \dot{x}, \dot{y}, \dot{z}})$ those of $v^a$, and 
${\sf v}^2 := \dot{\sf x}^2 + \dot{\sf y}^2 + \dot{\sf z}^2$. 

To construct the tripod's Lagrangian we apply Eq.~(\ref{Lagrangian2})
to the head and feet, and incorporate the constraints of
Eq.~(\ref{constraints}).\footnote{Because the coordinates are
  constructed from geodesic segments originating on $\gamma$, the
  struts responsible for enforcing the constraints are themselves very
  close to geodesic segments. They are not exactly geodesic, because
  the struts do not originate on $\gamma$ but at the tripod's head, a
  short distance away.} The tripod's position is described by the CM
variables $x$, $y$, and $z$, which are defined by Eq.~(\ref{CM_flat}),
with $\delta z$ assigned to be a solution to Eq.~(\ref{CMcorr}). We
recall from Sec.~\ref{sec:fermi} that $z$ represents the geodesic
distance between the CM and the reference observer, in the
longitudinal direction (increasing $R$), as measured in the observer's
rest frame. On the other hand, $x$ and $y$ measure the geodesic
distance between CM and observer in the transverse (angular)
directions. The CM is given the initial conditions 
$x = y = z = 0 = \dot{x} = \dot{y} = \dot{z}$, so that it is initially
moving with the reference observer. We wish to determine the CM's
motion at later times.    

We simplify the tripod's Lagrangian by removing all terms that 
do not involve the generalized coordinates and velocities (that is,
terms that are prescribed functions of $t$) and performing a
transformation to cylindrical coordinates $(\varrho, \varphi, z)$
defined by $x = \varrho \cos\varphi$ and $y = \varrho \sin\varphi$. 
The final expression is rather long, and we shall not display it here. 
Inspection of the Lagrangian reveals that it is independent of
$\varphi$, so that $p_\varphi = \partial L/\partial \varphi$ is a
constant of the motion. We also find that $p_\varphi \propto
\dot{\varphi}$, so that $p_\varphi = 0 = \dot{\varphi}$ at all times
by virtue of the initial conditions. We are therefore free to discard
all terms involving $\dot{\varphi}$ from the Lagrangian, and restrict
the phase space to $(\dot{\varrho}, \varrho, \dot{z}, z)$. 

To leading order in an expansion in powers of $v^2$, the
Lagrangian is given by 
\begin{equation} 
\bar{L} = \frac{1}{2} \bigl( \dot{z}^2 - B z^2 \bigr) 
+ \frac{1}{2} \bigl( \dot{\varrho}^2 - A \varrho^2 \bigr) 
+O(\epsilon),    
\end{equation} 
with $O(\epsilon)$ representing the relativistic corrections;  
$A$ and $B$ are given by Eq.~(\ref{ABCdef}). At this
order the equations of motion are  
\begin{equation} 
\ddot{z} = -B z + O(\epsilon), \qquad 
\ddot{\varrho} = -A \varrho + O(\epsilon), 
\end{equation} 
and these are recognized as components of the geodesic deviation
equation. The equations imply that with the initial conditions 
$z = \varrho = 0 = \dot{z} = \dot{\varrho}$, the solution shall be of
the form $z = O(\epsilon)$ and $\varrho = O(\epsilon)$, with
deviations from the reference geodesic coming entirely from the
relativistic terms in the Lagrangian. This fact allows us to simplify
the Lagrangian further, by eliminating terms that scale as
$\epsilon^3$ and higher powers of $\epsilon$. We thus obtain  
\begin{equation} 
\bar{L} = \frac{1}{2} \dot{z}^2 - \frac{1}{2} B z^2 
+ (a z + b \dot{z}) A + (c z + d \dot{z}) B + k \dot{z} 
+ O(\epsilon^3), 
\label{Lagrangian_final} 
\end{equation} 
with 
\begin{subequations} 
\label{abcd} 
\begin{align} 
a &:= \frac{m_0 m_1}{(m_0+3m_1)^2} \biggl[
\ell\sin\alpha \frac{d}{dt} (\ell\cos\alpha)
- \ell\cos\alpha \frac{d}{dt} (\ell\sin\alpha) \biggr] 
\frac{d}{dt} (\ell\sin\alpha), \\ 
b &:= \frac{m_0 m_1}{(m_0+3m_1)^2} \ell\sin\alpha \biggl[ 
\ell\cos\alpha \frac{d}{dt} (\ell\sin\alpha)
+ \frac{1}{2} \ell\sin\alpha \frac{d}{dt} (\ell\sin\alpha) \biggr], \\ 
c &:= -\delta z + \frac{3m_0 m_1}{2(m_0+3m_1)^2} \ell\cos\alpha 
\Biggl\{ \frac{m_0-3m_1}{m_0+3m_1} \biggl[ \frac{d}{dt}
    (\ell\cos\alpha) \biggr]^2
+ \biggl[ \frac{d}{dt} (\ell\sin\alpha) \biggr]^2 \Biggr\}, \\
d &:=  \frac{3m_0 m_1 (m_0-3m_1)}{2(m_0+3m_1)^3} 
(\ell\cos\alpha)^2 \frac{d}{dt} (\ell\cos\alpha), 
\end{align}
\end{subequations} 
and where $k$ is defined by Eq.~(\ref{CMcorr}). 
The terms $\frac{1}{2} (\dot{\varrho}^2 - A \varrho^2)$ were 
eliminated from the Lagrangian, because $\varrho$ and $\dot{\varrho}$
do not appear in the remaining terms of order $\epsilon^2$. This part
of the Lagrangian therefore decouples from the one displayed in
Eq.~(\ref{Lagrangian_final}), and its form implies that the solution
to the equations of motion with the stated initial conditions 
is $\varrho = O(\epsilon^3)$. The simplification has
therefore eliminated $\varrho$ and $\dot{\varrho}$ from the list of
dynamical variables, and the effective Lagrangian now depends solely
upon $z$ and $\dot{z}$. In this simplified description, the tripod's
internal motions produce a longitudinal displacement with respect to
the reference geodesic, but no transverse displacement.      

The equations of motion that follow from the Lagrangian of
Eq.~(\ref{Lagrangian_final}) can be cast in the first-order form  
\begin{equation} 
\dot{z} = p_z - bA - dB - k, \qquad 
\dot{p}_z = -B z + aA + cB,  
\label{EOM1} 
\end{equation} 
where $p_z$ is the momentum conjugate to $z$, or in the second-order
form  
\begin{equation} 
\ddot{z} + B z = F := (a-\dot{b}) A + (c-\dot{d}) B 
- b \dot{A} - d \dot{B}. 
\label{EOM2} 
\end{equation} 
As stated previously, the equations are to be integrated with the
initial conditions $z = 0 = \dot{z}$, so that the tripod begins
its journey on the reference geodesic.

\subsection{Tripod cycle} 

We consider two families of cycles for the tripod's internal
motions. The first is described by 
\begin{equation} 
\ell(t) = \frac{1}{2} \ell_0 (3 - \cos\alpha), \qquad 
\alpha(t) = \frac{2\pi t}{T} + \chi, 
\label{cycle2} 
\end{equation} 
in which $\ell$ oscillates and $\alpha$ is monotonic; $\chi$ is an
arbitrary phase constant and $T$ is the period. This cycle 
traces a cosine function in the $\alpha$--$\ell$ plane, and the area 
under the curve is equal to $3\pi\ell_0$. This family is particularly 
simple, and it allows us to integrate Eq.~(\ref{CMcorr}) analytically; 
we obtain 
\begin{align} 
\delta z &= \delta z_0 
+ \frac{3\pi^2 m_0 m_1 \ell_0^3}{160(m_0 + 3m_1)^3T^2} 
\Bigl[  (1320m_0 - 1980m_1) (\cos\alpha-\cos\chi) 
- (380m_0 - 570m_1) (\cos2\alpha-\cos2\chi) 
\nonumber \\ & \quad \mbox{} 
+ (40m_0+480m_1) (\cos3\alpha-\cos3\chi) 
- 405m_1 (\cos4\alpha-\cos4\chi) 
\nonumber \\ & \quad \mbox{} 
+ 108m_1 (\cos5\alpha-\cos5\chi) 
- 10m_1 (\cos6\alpha-\cos6\chi) \Bigr],   
\label{dz_cycle2} 
\end{align} 
where $\delta z_0 = \delta z(0)$ is a constant of
integration. Inserting these expressions in Eqs.~(\ref{abcd}) returns  
\begin{subequations} 
\label{abcd_cycle2} 
\begin{align}  
a &= \frac{\pi^2 m_0 m_1 \ell_0^3}{8 (m_0+3m_1)^2 T^2} \Bigl[ 37  
- 129\cos\alpha + 74\cos2\alpha - 15\cos3\alpha + \cos4\alpha \Bigr],
\\ 
b &= -\frac{\pi m_0 m_1 \ell_0^3}{128 (m_0+3m_1)^2T^2} \Bigl[
126\sin\alpha - \sin2\alpha - 351\sin3\alpha + 220\sin4\alpha 
- 45\sin5\alpha + 3\sin6\alpha \Bigr], 
\\ 
c &= -\delta z_0 
-\frac{3\pi^2 m_0 m_1 \ell_0^3}{160 (m_0+3m_1)^3T^2} \Bigl[ 
(560m_0 + 270m_1) - (1320m_0 - 1980m_1) \cos\chi 
+ (380m_0 - 570m_1) \cos2\chi 
\nonumber \\ & \quad \mbox{} 
- (40m_0 + 480m_1) \cos3\chi 
+ 405m_1 \cos4\chi - 108m_1 \cos5\chi + 10m_1 \cos6\chi  
- (60m_0 + 3780m_1)\cos\alpha 
\nonumber \\ & \quad \mbox{} 
+ (180m_0 + 2220m_1)\cos2\alpha 
- (20m_0 + 1680m_1)\cos3\alpha 
+ 1005m_1 \cos4\alpha 
- 252m_1 \cos5\alpha + 20m_1 \cos6\alpha \Bigr], 
\\ 
d &= -\frac{3\pi m_0 m_1 (m_0-3m_1) \ell_0^3}{128 (m_0+3m_1)^3T^2} 
\Bigl[ 138\sin\alpha - 149\sin2\alpha + 153\sin3\alpha - 76\sin4\alpha 
+ 15\sin5\alpha - \sin6\alpha \Bigr]. 
\end{align} 
\end{subequations} 

The second family of cycles is described by 
\begin{equation} 
\ell(t) = \frac{1}{2} \ell_0 \bigl[ 3 - \cos(2\pi t/T + \chi) \bigr]\, \qquad  
\alpha(t) = \frac{\pi}{12} \bigl[ 3 - \cos(2 \pi t/T) \bigr], 
\label{cycle1} 
\end{equation} 
in which $\ell$ oscillates between $\ell_0$ and $2\ell_0$ in the
course of a complete period, while $\alpha$ oscillates between $\pi/6$
and $\pi/3$. This cycle traces an ellipse in the $\alpha$--$\ell$ plane, 
sweeping out a surface with area $(\pi^2/24) \ell_0 \sin\chi$.

We shall focus our attention mostly on the cycle of
Eq.~(\ref{cycle2}), and take advantage of its simplicity. We shall
also, however, present numerical results for the cycle of
Eq.~(\ref{cycle1}).   

\section{Tripod motion: Results}  
\label{sec:tripod2}

\subsection{de Sitter spacetime}
\label{subsec:deSitter} 

We begin with a discussion of the tripod's motion in de Sitter
spacetime, for which $f = 1-k^2 R^2$ and the curvatures of
Eq.~(\ref{ABCdef}) are $A = B = -4/(9 t_0^2)$, where $t_0 := 2/(3k)$
is a conveniently defined cosmological time scale. The solution to
Eq.~(\ref{EOM2}) with vanishing initial conditions is
\begin{equation} \label{zdS}
z(t) = \frac{2}{3 t_0} \int_0^t{g(t') \sinh[2(t'-t)/3t_0] dt'},
\end{equation}
where $g := a - \dot{b} + c - \dot{d}$. From Eq.~(\ref{abcd}) we see
that when $\ell$ and $\alpha$ are periodic, $g(t)$ will also be a
periodic function of $t$ with mean 
$g_0 := T^{-1} \int_0^T{g(t)dt} =
T^{-1} \int_0^T{[a(t)+c(t)]dt}$. In particular, for the cycle of 
Eq.~(\ref{cycle2}),
\begin{align}
g_0 &= -\delta z_0 + \frac{\pi^2 m_0 m_1 \ell_0^3}{160 (m_0 + 3 m_1)^3 T^2} 
 \Bigl[ -470(2 m_0 - 3 m_1) + 1980 (2 m_0 -3 m_1) \cos\chi 
- 570 (2 m_0 - 3 m_1) \cos 2\chi 
\nonumber \\ & \quad \mbox{} 
+120 (m_0 + 12 m_1) \cos 3\chi
-1215 m_1 \cos 4\chi + 324 m_1 \cos 5 \chi - 30 m_1 \cos 6 \chi \Bigr].
\end{align}
When $T \ll t_0$, which defines the regime of rapid cycles, the
oscillatory terms in $g(t)$ give a negligible contribution to the
integral of Eq.~(\ref{zdS}), and $z(t)$ is well approximated by 
\begin{equation} 
z(t) = g_0 \bigl[ 1 - \cosh(2t/3t_0) \bigr].
\label{z_deSitter} 
\end{equation}  
Now, for an arbitrary choice of $\delta z_0$, $g_0$ is nonzero and the
geodesic observer sees the tripod drifting away exponentially. Such a
drift is paradoxical, because de Sitter spacetime is maximally
symmetric, and there can be no preferred direction for the tripod's
motion relative to the reference geodesic. But we have the option to
eliminate this drift by adjusting $\delta z_0$ so that  
$g_0 =0$. For this specific choice of CM we have that $z(t) = 0$,
which means that when the CM is placed initially on a geodesic in de
Sitter spacetime, it will continue to move on this geodesic (at least
in the regime of rapid cycles). This choice of CM gives us the
expected geodesic motion for a tripod in de Sitter spacetime, and 
therefore provides us with a sensible prescription for the
determination of $\delta z_0$. This prescription can be applied to any
spacetime. The point remains, however, that geodesic motion reflects a
choice of CM, and that an alternative choice would generically produce
the drift described by Eq.~(\ref{z_deSitter}).   

\subsection{Schwarzschild spacetime}

We next turn to Schwarzschild spacetime, for which $f = 1 - 2GM/R$; 
the tripod falls toward a spherical body of mass $M$. For $\gamma$ we
choose a marginally bound, radial geodesic with $E = 1$, so that 
$dR/dt = -\sqrt{1-f} = -\sqrt{2GM/R}$, which integrates to 
\begin{equation} 
R(t) = \biggl[ \frac{9}{2} G M (t_0 - t)^2 \biggr]^{1/3}, 
\label{R(t)} 
\end{equation} 
where $t_0$ is now the time at which $R$ is formally equal to zero; recall
that in our notation, $t$ is proper time on $\gamma$. The curvatures
of Eq.~(\ref{ABCdef}) become  
\begin{equation} 
A = \frac{2}{9(t_0-t)^2}, \qquad 
B = -\frac{4}{9(t_0-t)^2} 
\label{AB} 
\end{equation} 
for this choice of spacetime and reference geodesic. 

With $B$ given by Eq.~(\ref{AB}), the solution to Eq.~(\ref{EOM2}) with vanishing
initial conditions is  
\begin{equation} \label{zS}
z (t) = \frac{3}{5} (t_0-t)^{-1/3} \int_0^t (t_0-t')^{4/3} F(t')\, dt'
- \frac{3}{5} (t_0-t)^{4/3} \int_0^t (t_0-t')^{-1/3} F(t')\, dt'.
\end{equation} 
To evaluate the integrals we discard the terms involving $\dot{A}$ and
$\dot{B}$ in $F$, which give negligible contributions until $t$
approaches $t_0$. This gives $F \simeq 2 \tilde{g}/[9(t_0-t)^2]$, with
$\tilde{g} := a - \dot{b} - 2(c - \dot{d})$. The function
$\tilde{g}(t)$ is periodic with mean 
$\tilde{g}_0 := T^{-1} \int_0^T{\tilde{g}(t)dt} 
= T^{-1} \int_0^T{[a(t)- 2 c(t)]dt}$, and for the cycle described by
Eq.~(\ref{cycle2}), 
\begin{align} 
\tilde{g}_0 &:= -\delta z_0 
+ \frac{\pi^2 m_0 m_1 \ell_0^3}{80(m_0+3m_1)^3} \Bigl[ 
(2050m_0+1920m_1) - (3960m_0-5940m_1)\cos\chi
+ (1140m_0-1710m_1)\cos 2\chi
\nonumber \\ & \quad \mbox{} 
- (120m_0+1440m_1)\cos 3\chi
+ 1215m_1 \cos 4\chi
- 324m_1 \cos 5\chi 
+30m_1 \cos 6\chi \Bigr].
\label{g0} 
\end{align} 
Discarding the oscillations in $\tilde{g}(t)$, because they give 
negligible contributions to the integrals in the regime of rapid
cycles,  and keeping only the mean, Eq.~(\ref{zS}) becomes
\begin{equation} 
z(t) = -\frac{2}{5} \tilde{g}_0 \bigl[ 1 - (1-t/t_0)^{-1/3} \bigr] 
- \frac{1}{10} \tilde{g}_0 \bigl[ 1 - (1-t/t_0)^{4/3} \bigr].
\label{z_approx} 
\end{equation} 
When $t/t_0$ is small, Eq.~(\ref{z_approx}) reduces to $z \sim
\frac{1}{9} \tilde{g}_0 (t/t_0)^2$. On the other hand, 
\begin{equation} 
z \sim \frac{2}{5} \tilde{g}_0 (1 - t/t_0)^{-1/3}
\label{z_asymp} 
\end{equation} 
when $t$ approaches $t_0$. 

Equation (\ref{z_approx}) describes a drift relative to the reference
geodesic, and the drift is proportional to $\tilde{g}_0$ given by
Eq.~(\ref{g0}). Because $\tilde{g}_0$ depends on $\delta z_0$, we
see once again that the description of the tripod's motion depends 
sensitively on the choice of CM. An option would be to adjust 
$\delta z_0$ so that $\tilde{g}_0 = 0$, and to eliminate the
drift in Schwarzschild spacetime. This would make an alternative
prescription for the complete determination of the CM, and adopting it
for de Sitter spacetime would give $g_0 \neq 0$ and a drift relative
to the reference geodesic; we would recover the same paradox as
described in Sec.~\ref{subsec:deSitter}. A more sensible prescription
is the one adopted in Sec.~\ref{subsec:deSitter}, which gives no drift
in de Sitter spacetime. With this original prescription, and for the
cycle of Eq.~(\ref{cycle2}), we have that   
\begin{equation}
\tilde{g}_0 = \frac{111 \pi^2 m_0 m_1 \ell_0^3}
{8(m_0 + 3 m_1)^2 T^2}, 
\label{g0tilde} 
\end{equation}
and there is a drift in Schwarzschild spacetime.  

\begin{figure} 
\includegraphics[width=0.6\linewidth]{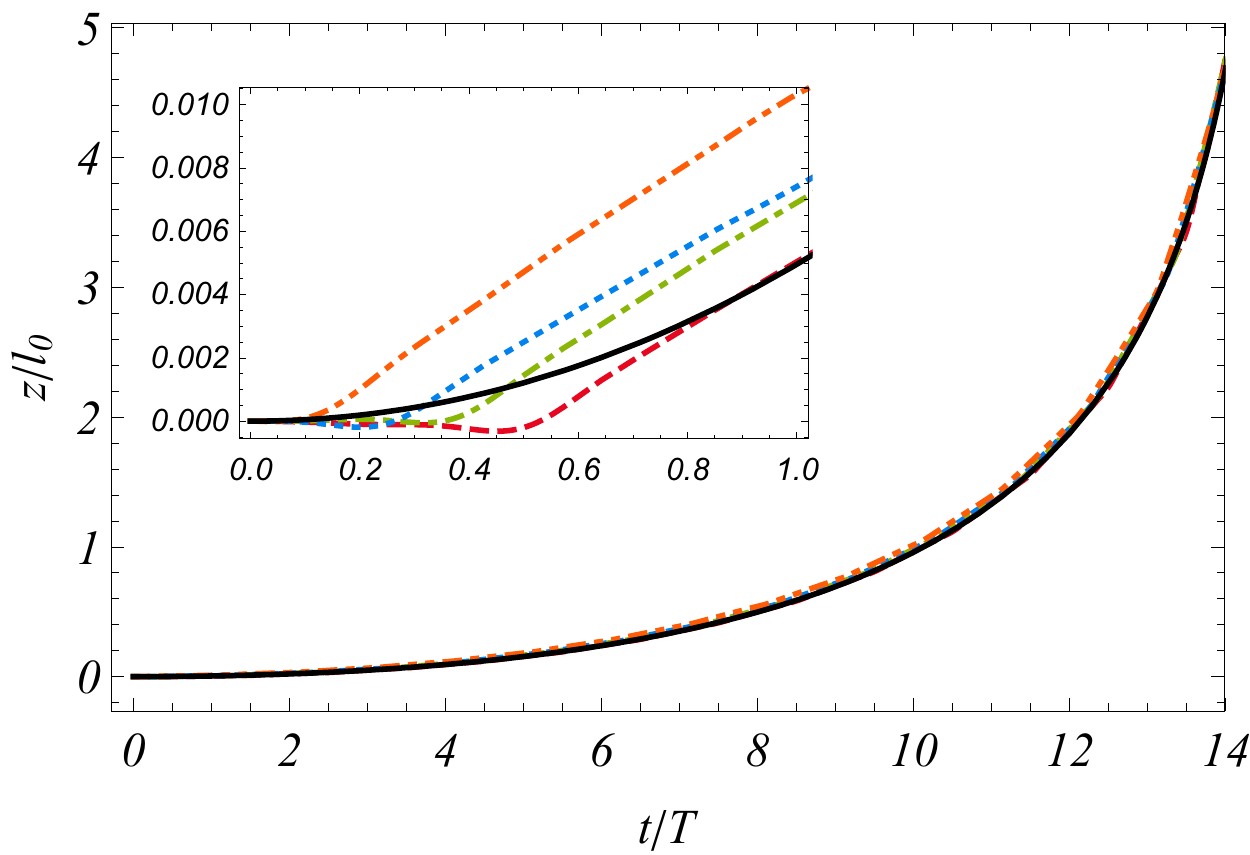}
\caption{Center of mass drift in Schwarzschild spacetime for the cycle
  of Eq.~(\ref{cycle2}), with $t_0/T = 15$, $m_0/(m_0+3m_1) = 0.7$, 
  and $m_1/(m_0+3m_1) = 0.1$. The vertical axis represents $z$, the
  tripod's longitudinal displacement with respect to the reference
  geodesic, measured in units of $\ell_0$. The horizontal axis
  represents $t$, measured in units of the cycle period $T$. The phase 
  parameter is set to $\chi = 0$ (red curves, dash), 
  $\chi = \pi/4$ (green curves, dash and single dot), 
  $\chi = \pi/2$ (blue curves, dot), and 
  $\chi = 3\pi/4$ (orange curves, dash and double dots). The broken, 
  colored curves are the result of numerical integrations, while the
  solid, black curves are the analytical approximation of 
  Eqs.~(\ref{z_approx}) and (\ref{g0tilde}). A positive $z$ indicates
  that the tripod drifts upward, in the direction of increasing $R$ in 
  the spacetime.}    
\label{fig:cycle2} 
\end{figure}

\begin{figure} 
\includegraphics[width=0.6\linewidth]{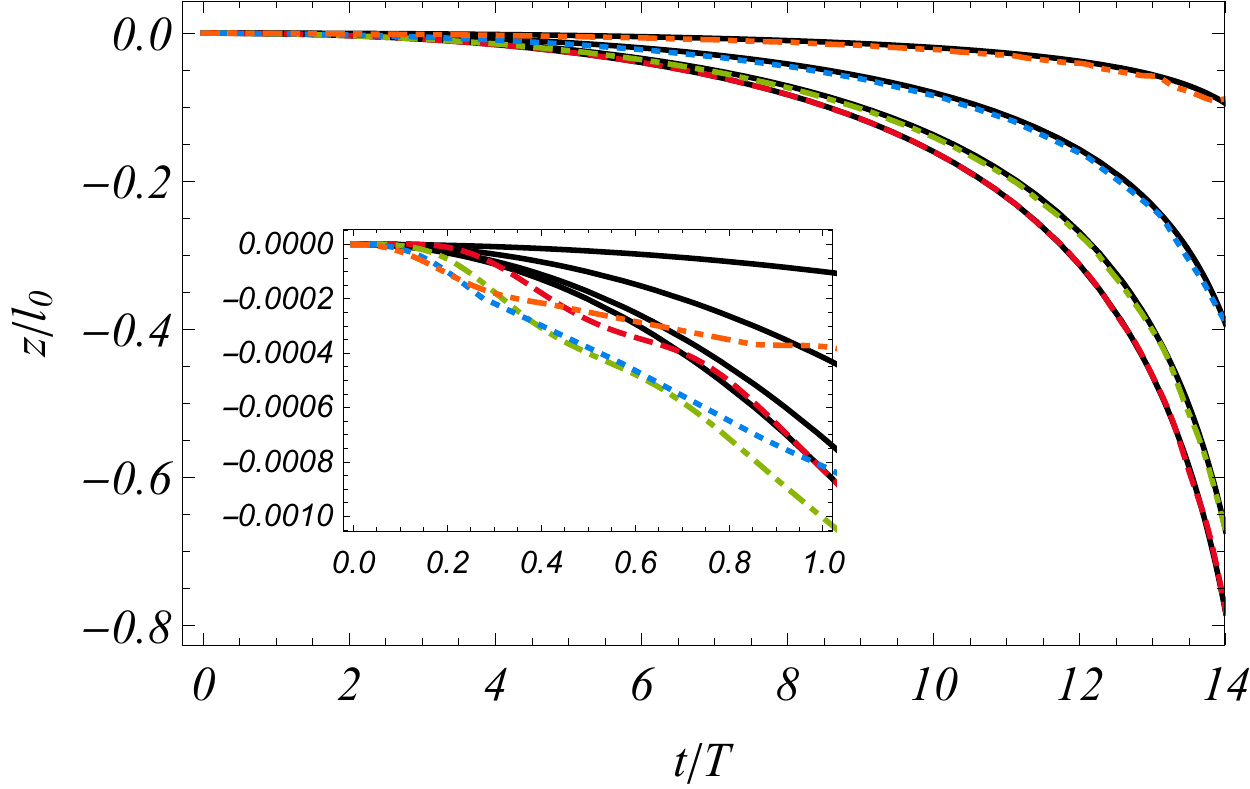}
\caption{Same as Fig.~\ref{fig:cycle2} for the cycle of
  Eq.~(\ref{cycle1}). Again we set $t_0/T = 15$, $m_0/(m_0+3m_1) =
  0.7$, and $m_1/(m_0+3m_1) = 0.1$. The phase parameter is set to 
  $\chi = 0$ (red curves, dash), 
  $\chi = \pi/4$ (green curves, dash and single dot), 
  $\chi = \pi/2$ (blue curves, dot), and 
  $\chi = 3\pi/4$ (orange curves, dash and double dots). The broken,
  colored curves are the result of numerical integrations, while the
  solid, black curves are the analytical approximation of
  Eq.~(\ref{z_approx}), with the appropriate $\tilde{g}_0$ for this
  cycle evaluated at the corresponding $\chi$ ($\tilde{g}_0$ depends
  on $\chi$ for this cycle). A negative $z$ indicates that the tripod
  drifts downwards, in the direction of decreasing $R$ in the spacetime.}  
\label{fig:cycle1} 
\end{figure}

In Fig.~\ref{fig:cycle2} we show that Eq.~(\ref{z_approx}) with the 
$\tilde{g}_0$ of Eq.~(\ref{g0tilde}) is indeed a good approximation to
the exact solution of Eq.~(\ref{EOM2}) with initial conditions 
$z = 0 = \dot{z}$ at $t=0$, for the cycle of Eq.~(\ref{cycle2}) and a 
$\delta z_0$ determined by the de Sitter condition $g_0 = 0$. A
similar comparison for the cycle of Eq.~(\ref{cycle1}) is presented in 
Fig.~\ref{fig:cycle1}. We note that whether the CM lags behind or
sprints forward relative to the geodesic observer depends on the
specific choice of cyclic motion. 

The analytical expression of Eqs.~(\ref{z_approx}) and (\ref{g0tilde})
allows us to calculate the tripod's drift after one cycle of its
internal motions. This is measured by $z(T)$, which evaluates to 
$\frac{1}{9} \tilde{g}_0 (T/t_0)^2$ after taking into account that $t_0 \gg T$.   
Substituting Eq.~(\ref{g0}) and relating $t_0^{-2}$ to the spacetime
curvature through Eqs.~(\ref{R(t)}) and (\ref{AB}), we write this as 
\begin{equation} 
z(T) = \frac{37}{9\pi} \frac{m_0 m_1}{(m_0 + 3m_1)^2}\, \ell_0^2\, 
(3\pi \ell_0)\, \frac{GM}{R^3}.
\end{equation} 
We observe that the drift scales with $m_0 m_1/(m_0 + 3m_1)^2$, $\ell_0^2$, 
the area $3\pi \ell_0$ of the $\alpha$--$\ell$ plane swept out by the 
tripod's cycle, and with the curvature $GM/R^3$. All these ingredients
are also featured in Wisdom's Eq.~(20) \cite{wisdom:03}, and we
therefore have recovered the essence of his result. We do not get a
precise match --- the numerical prefactor is different --- because we
adopt a different cycle for the tripod's internal motions, model the
struts in a slightly different way, and work in the Fermi frame of the
free-falling observer instead of the static frame of the Schwarzschild
spacetime.  

As a final comment, it is interesting to note that the asymptotic relation 
(\ref{z_asymp}), as $t$ approaches $t_0$, describes a geodesic of 
the Schwarzschild spacetime. Indeed, the homogeneous version of
Eq.~(\ref{EOM2}), $\ddot{z} + Bz = 0$, is a component of the geodesic
deviation equation, and its general solution 
\begin{equation} 
z_{\rm geo}(t) = 
\frac{1}{5} \bigl[ 4z(0) + 3 t_0 \dot{z}(0)\bigr] (1-t/t_0)^{-1/3} 
+ \frac{1}{5} \bigl[ z(0) - 3 t_0 \dot{z}(0) \bigr] (1-t/t_0)^{4/3}  
\end{equation} 
describes a geodesic of the Schwarzschild spacetime that neighbors the
reference geodesic $\gamma$. The first term dominates as $t$
approaches $t_0$, and we see that with 
\begin{equation} 
4z(0) + 3 t_0 \dot{z}(0) = 2 \tilde{g}_0, 
\label{geod_IC} 
\end{equation} 
$z_{\rm geo}$ becomes asymptotically equal to $z$. The world line 
described by Eq.~(\ref{z_asymp}), therefore, is a geodesic with initial
conditions constrained by Eq.~(\ref{geod_IC}). 
This allows us to give an interpretation to the numerical results of 
Figs.~\ref{fig:cycle2} and \ref{fig:cycle1}. 
What we see is the tripod's CM gradually transiting to a new geodesic
described by Eq.~(\ref{z_asymp}) after being launched from the
reference geodesic. The tripod is initially directed along $\gamma$,
but the coupling between its internal motions and the spacetime
curvature prevents it from following the reference geodesic. The
motion is therefore nongeodesic for $0 < t < t_0$, but it becomes
increasingly geodesic as $t \to t_0$, that is, as the tripod
approaches the curvature singularity at $R = 0$. The approach to the
singularity follows a geodesic, irrespective of the internal
motions\footnote{We should note that the description of the motion
refers to the Fermi frame introduced in Sec.~\ref{sec:fermi}. The metric
is approximate, and its domain of validity becomes increasingly
narrow as the curvature increases. The approach to the singularity
must therefore be handled with care; $t$ cannot be allowed to be too
close to $t_0$.}. The same reasoning can be applied to de Sitter
spacetime, when a CM definition such that $g_0 \neq 0$ is adopted.  
In this case, the motion asymptotes to $z \sim \frac{1}{2} g_0 e^{2t/3t_0}$ 
when $t \gg t_0$, and this describes a geodesic of de Sitter spacetime 
with initial conditions constrained by $2 z(0) + 3t_0 \dot{z}(0) = 2g_0$. 

\subsection{Conclusion} 

Our main conclusion in this section is that the motion of the tripod
relative to the reference geodesic depends sensitively on the choice
of internal cycle, but also on the choice of $\delta z_0$, the initial
relativistic adjustment to the CM. The prescription advanced in
Sec.~\ref{subsec:flat} left this quantity 
undetermined, and the prescription must therefore be completed by a
choice of $\delta z_0$. A sensible (though not unique) option is to
choose $\delta z_0$ so that $g_0 = 0$, thereby ensuring that a tripod
placed in de Sitter spacetime does not drift away from the initial
geodesic; this choice is motivated by the absence of a preferred
direction in a maximally symmetric spacetime. Adopting the same CM
when the tripod is placed in the Schwarzschild spacetime gives rise to
a drift, with an approximate description given by
Eqs.~(\ref{z_approx}) and (\ref{g0tilde}), in accordance to Wisdom's
original results \cite{wisdom:03}. 

\section{Reconciliation with the Mathisson-Papapetrou-Dixon equations} 
\label{sec:MPD} 

The Mathisson-Papapetrou-Dixon (MPD) equations, 
\begin{equation} 
\frac{Dp_\alpha}{d\tau} = \frac{1}{2} S^{\mu\nu} u^\lambda
R_{\mu\nu\lambda\alpha}, \qquad 
\frac{DS_{\alpha\beta}}{d\tau} = 2 p_{[\alpha} u_{\beta]},  
\label{MPD} 
\end{equation}  
are meant to govern the motion of a generic extended body, in a
pole-dipole approximation that neglects higher multipole moments 
but keeps the momentum vector $p_\alpha$ and spin tensor
$S_{\alpha\beta}$; the velocity vector $u^\alpha$ is tangent to the
world line, and $D/d\tau$ denotes a covariant derivative with respect
to proper time $\tau$. Many derivations of these equations have been
provided (see, for example, Refs.~\cite{mathisson:10, papapetrou:51a,
  bailey-israel:75}), with the most comprehensive analysis supplied by
Dixon \cite{dixon:70a, dixon:70b, dixon:74}. The puzzle
that concerns us in this section is that while the MPD equations would 
be thought to adequately govern the motion of the tripod, the detailed
examination carried out by Silva, Matsas, and Vanzella
\cite{silva-matsas-vanzella:16} indicates that Eqs.~(\ref{MPD}) seem 
to be incompatible with the swimming motion described in
Sec.~\ref{sec:tripod2}.  
 
One reason to suspect that Eqs.~(\ref{MPD}) may not apply to the
tripod is that their derivation relies heavily on conservation of
energy-momentum, as embodied by $\nabla_\beta T^{\alpha\beta} = 0$,
where $T^{\alpha\beta}$ is the energy-momentum tensor of the extended 
body. (Alternative derivations depend on the existence of a Lagrangian
for the extended body, and this automatically enforces energy-momentum
conservation.) The tripod, on the other hand, is a constrained mechanical
system that does not conserve energy and momentum: the external agents
responsible for the internal motions must supply energy and momentum
to keep the tripod on its cycle.   

This issue can be investigated by generalizing the Lagrangian-based
derivation of Eqs.~(\ref{MPD}) provided by Bailey and 
Israel \cite{bailey-israel:80} to a constrained system. To showcase
the modifications that result from the introduction of constraints, we 
consider the illustrative case of two particles moving freely in
spacetime, except for a holonomic constraint that keeps their
separation equal to a prescribed vector field $f^\alpha$. This is a
covariant formulation of the type of tripod model introduced in
Sec.~\ref{sec:tripod1}, and generalization to any number of particles
(like four, for an actual tripod) is immediate.  

Bailey and Israel place the two particles on physical world lines
${\cal P}_1$ and ${\cal P}_2$, but describe their motion in terms of
a reference world line $\cal C$ that represents an arbitrary choice 
of ``center of mass''. The world lines are given an arbitrary
parameter $t$, the tangent vector to ${\cal C}$ is $u^\alpha$, and the 
separation between $\cal C$ and ${\cal P}_A$ at fixed $t$ is 
$-\sigma^\alpha_A$, with $A = 1, 2$ labelling the particle. The
Lagrangian $L_A$ of each particle is the usual $-m_A d\tau_A/dt$
evaluated on ${\cal P}_A$, which is rewritten in terms of fields on
${\cal C}$; an explicit expression for the Lagrangian is given by
their Eq.~(76). The complete Lagrangian for the constrained system is 
$L = L_1 + L_2 + L_{\rm cons}$, with
\begin{equation} 
L_{\rm cons} = \lambda_\alpha \bigl( \sigma^\alpha_2 
- \sigma^\alpha_1 + f^\alpha \bigr),  
\end{equation} 
where $\lambda_\alpha(t)$ is a Lagrange multiplier, and
$f^\alpha(t)$ is the prescribed separation between the
particles. The dynamical variables are defined by 
\begin{equation} 
p_\alpha := \frac{\partial L}{\partial u^\alpha}, \qquad 
S_{\alpha\beta} := 
2 \sigma_{1 [\alpha} \frac{\partial L}{\partial \dot{\sigma}_1^{\beta]}}  
+ 2 \sigma_{2 [\alpha} \frac{\partial L}{\partial \dot{\sigma}_2^{\beta]}},
\end{equation} 
where $\dot{\sigma}^\alpha_A := D \sigma_A^\alpha/dt$ is the 
covariant derivative of $\sigma^\alpha_A$ along ${\cal C}$. 

Equations of motion for $p_\alpha$ and $S_{\alpha\beta}$ follow from
the requirement that $S := \int L\, dt$ be stationary under arbitrary
variations of ${\cal P}_A$ and ${\cal C}$; the derivation also
involves an identity deduced from the invariance of the action under
an infinitesimal coordinate transformation. Variation with respect to
${\cal P}_A$ implicates the constraints, and Eq.~(50) from 
\cite{bailey-israel:80} generalizes to 
\begin{equation} 
\frac{D}{dt} \frac{\partial L_1}{\partial \dot{\sigma}^A_1} 
- \frac{\partial L_1}{\partial \sigma^\alpha_1} = -\lambda_\alpha,
\qquad 
\frac{D}{dt} \frac{\partial L_2}{\partial \dot{\sigma}^A_2} 
- \frac{\partial L_2}{\partial \sigma^\alpha_2} = +\lambda_\alpha; 
\end{equation} 
these equations can be used to determine the Lagrange
multiplier. Variation with respect to ${\cal C}$ reproduces 
Eq.~(47) from \cite{bailey-israel:80} without change (we set the 
electric charge $e$ to zero), and their Eq.~(51) acquires new terms
coming from $L_{\rm cons}$.  

Putting all this together, truncating the description of the motion to 
the pole-dipole approximation, and setting $t = \tau = \mbox{proper
  time on $\cal C$}$, we find that the equations of motion take the
form of  
\begin{equation} 
\frac{Dp_\alpha}{d\tau} = \frac{1}{2} S^{\mu\nu} u^\lambda
R_{\mu\nu\lambda\alpha}, \qquad 
\frac{DS_{\alpha\beta}}{d\tau} = 2 p_{[\alpha} u_{\beta]} 
+ 2 \lambda_{[\alpha} f_{\beta]},  
\label{MPD_mod} 
\end{equation}  
which differ from Eqs.~(\ref{MPD}) by an additional torque term
provided by the constraints. As expected, the original MPD equations
do not apply to a constrained Lagrangian system.   

Because Eqs.~(\ref{MPD_mod}) are derived from a Lagrangian, they must
be physically equivalent (after generalization to four particles, and
approximation to weakly relativistic motion) to the equations of
motion obtained in Sec.~\ref{sec:tripod1}, which also follow from a
Lagrangian. To establish the equivalence explicitly would be 
difficult, because the two sets of equations implicate different
variables, and because the constraints are implemented differently in
each approach: in Sec.~\ref{sec:tripod1} the constraints were solved
to express the Lagrangian in terms of the CM variables, while in this
section they are enforced with Lagrange multipliers. Another 
obstacle toward establishing the equivalence of the two formulations
is that Eqs.~(\ref{MPD_mod}) remain empty of content until a relation
between $p_\alpha$ and $u_\alpha$ is specified through the selection 
of a suitable ``center of mass''; as we saw back in
Sec.~\ref{sec:tripod1}, this can be a delicate matter. But in spite of
these obstacles, we can be confident that Eqs.~(\ref{EOM1}) or
(\ref{EOM2}) and (\ref{MPD_mod}) describe the same system, because the
two sets of equations originate from the same Lagrangian. In this
admittedly incomplete way, we can reconcile Wisdom's swimming with the 
MPD framework.   

It seems to us that the selection of a ``center of mass'' is probably
the most critical aspect of the comparison between Wisdom's results 
\cite{wisdom:03} and the MPD framework; the additional terms in
Eq.~(\ref{MPD_mod}) may well be incidental. Indeed, we could imagine
formulating a complete tripod model that includes all the springs and
rubber bands that are dynamically responsible for the internal
motions. Such a model could be described in terms of a Lagrangian or a
conserved energy-momentum tensor, and such a tripod would be expected
to satisfy Eqs.~(\ref{MPD}). But the selection of a ``center of mass''
for this model would be a delicate affair, and the precise relation
between $p_\alpha$ and $u_\alpha$ might be complicated by the many
details of the tripod's design. For example, the choice of ``center of
mass'' that comes with the oft-used covariant spin supplementary
condition, $p_\alpha S^{\alpha\beta} = 0$, might be entirely
inadequate for such an object. In such circumstances, the relation
between Eqs.~(\ref{MPD}) and the actual motion of the tripod might be
very subtle and difficult to describe. This is another path of
reconciliation.  

\bibliography{/Users/poisson/writing/papers/tex/bib/master} 

\end{document}